# Toward Better Physics Labs for Future Biologists

K. Moore[1*], J. Giannini[2] & W. Losert[1,2*]

[1] *Department of Physics, University of Maryland, College Park, MD 20742*
[2] *Biophysics Program, University of Maryland, College Park, MD 20742*

**Abstract.** We have developed a set of laboratories and hands on activities to accompany a new two-semester interdisciplinary physics course that has been successfully developed and tested in two small test classes of students at the University of Maryland, College Park (UMD) in 2012-2013. We have designed the laboratories to be taken accompanying a reformed course in the student's second year, with calculus, biology, and chemistry as prerequisites. This permits the laboratories to include significant content on physics relevant to cellular scales, from chemical interactions to random motion and charge screening in fluids. We also introduce the students to research-grade equipment and modern physics analysis tools in contexts relevant to biology, while maintaining the pedagogically valuable open-ended laboratory structure of reformed laboratories. Preliminary student results from these two small test classes are discussed.

## I. INTRODUCTION

The increasingly urgent calls for changes to the undergraduate introductory physics curriculum for life sciences majors (IPLS)[1,2,3,4,5] have spurred much activity in the physics community. The physics education community has been re-examining our current curricular methods and choices, and biological physicists have been working to make significant changes to the physics content covered in introductory classes, with both communities striving to provide a better physics foundation for aspiring life scientists.

The NEXUS/Physics project,[6] described in detail elsewhere in this volume,[7] has brought together physics education researchers and biological physicists to tackle a reform of both physics content and pedagogy, with significant input and contributions from the biomedical community. As part of this project, we have endeavored to create a new IPLS laboratory curriculum focused on physics relevant to living systems that blends epistemic commitments[8] with authentic biological contexts[7,9] as a first step toward better physics labs for budding biologists.

Our laboratory development team, which includes both biological physicists and physics education researchers, zeroed in on two main aspects of the physics foundation for future life scientists that stand out as particularly relevant to new laboratories. First, from a physics education research perspective, physics provides a platform to develop modeling skills, the ability to frame observations in the context of models, and in particular mathematical, quantitative models, as we will describe in detail later. This is a well-recognized foundation where well-designed laboratories may make a significant contribution. Indeed quantitative modeling, which traditionally had been used in only a few areas of biology such as ecology and neuroscience, is now becoming part of virtually all biological and biomedical research, driven by a rapid surge in quantitative biological information.

Second, from a biological physics perspective, the physics needed to gain insights into living systems is different from the physics taught in many introductory courses.[10,11] Physics at the cellular scale, for example, involves Brownian motion and entropy, as well as fluid flow, ionic charges, and charge screening—topics not usually covered in physics classes for life scientists. Similarly, the physics topics relevant at the scale of organisms or populations, such as scaling relations, are not usually the focus of introductory classes or labs. Many of the physical principles relevant to living systems, especially physics relevant to nanoscopic systems such as proteins and DNA, and microscopic systems such as cells, are very distinct from the physics of macroscopic objects that students encounter in everyday life. Thus the introductory physics laboratories provide the unique opportunity to build from scratch hands-on experience with forces and motion at the nanoscopic and microscopic scales.

Before describing our proposed IPLS lab curriculum, it is worthwhile to consider the styles of curricula currently available. At most institutions, introductory physics students are given either a "traditional" or a "reform" laboratory curriculum. While these curricula have their advantages, we argue that they are inadequate when facing the challenge of educating this new generation of life sciences students.

"Traditional" laboratory curricula, of the explicitly-directed, "cookbook" variety, have the advantages of covering a wide range of physics content and employing analysis methods that can be quite sophisticated. These labs are often used to demonstrate theoretical principles already presented in the lecture portion of the course. The disadvantages directly resulting from the breadth of topic coverage and the intricate mathematical analysis are that these labs require heavy-handed guidance that often has the down-side of stifling student attempts to engage in sense-making.[12] Students often view these types of labs as a sequence of disconnected steps that must be followed, which have no relation to each other or to the rest of the course.[8] Moreover, the physics content covered by these labs, intended to lay the groundwork for physics investigations in advanced physics courses, fails to meet the needs of life sciences students, for whom the introductory physics class is likely to be the only physics course they take. These traditional labs may also employ lab techniques and tools that are hopelessly outdated: ticker-tape and stopwatches are not the tools of modern science.

In recognition of the limitations of traditional lab curricula, several "reform" lab curricula have been created, e.g. UMD's Scientific Community Labs (SCL)[13,14] and Rutgers' Investigative Science Learning Environment (ISLE)[15,16,17] labs. These lab curricula share an emphasis on developing scientific reasoning through experimental design and thoughtful data and error analysis, with an explicit focus on epistemological considerations: what it means to "know" and what counts as evidence. Through explicit engagement in sense-making, group work, and communication, these reformed labs show dramatic gains in student reasoning and scientific processing.[18,19,20] Yet the physics contexts of these labs still reflects traditional content selection and employs the same kinds of low-technology, outdated equipment. While these reformed labs succeed in demonstrating epistemic gains, they are not successful in helping students see how the physics labs can help prepare them for their future careers.[21] Thus we argue that a new kind of lab curriculum needs to be developed for IPLS courses.

## II. OUR VISION FOR A NEW IPLS LAB CURRICULUM

We aim to retain the epistemic gains sought by the various reformed lab curricula while seeking changes that particularly benefit our target student demographic:

- a focus on physics relevant to microscopic and living systems;
- the use of 21st century tools and software;
- the ability to engage with data-rich environments; and
- preparation for future contributions to biomedical research.

Presenting the students with physics relevant to microscopic and living systems allows us to engage their interest and build a stronger conceptual understanding. By accessing, challenging, and revising their grasp of scientific concepts learned in previous biology and chemistry classes, it is possible to forge a stronger connection than would be formed by presenting traditional physics topics. For example, students engage in investigations of randomness and Brownian motion, charges and motion in fluids, and fluorescence. Though they have encountered many of these topics in their previous biology and chemistry courses, their prior knowledge often reflects a cursory exposure with limited depth of understanding.

In using 21st century tools and software, we hope to convey to students that physics is a modern science. We also aim to show the applicability of physical measurement in analyzing biological processes and to provide students with modern analysis tools that will enable them to engage with the data-rich environments they will encounter in their research. The use of modern equipment such as video capture (through digital cameras in the microscopes) and real-time data collection probes (like the spectrometer) generate a wealth of data that most students find overwhelming. By presenting students with sophisticated analysis methods for large(r) datasets (such as $\log(r^2)$ vs. $\log(t)$ plots, discussed below) and with flexible software packages (such as NIH's ImageJ for image and video analysis), students have the opportunity to learn cutting-edge skills with immediate applicability in biomedical research.

In addition to these benefits and the lab skills called for in Competency E2 of the HHMI-NEXUS project (i.e., *Demonstrate understanding of the process of scientific inquiry, and explain how scientific knowledge is discovered and validated*),[3] the development of the lab curriculum focuses on five main threads: modeling; experimental design/protocol development; error analysis; technical lab skills; and interdisciplinary thinking.

### A. Modeling

Crucial to a scientific mode of thinking and to the development of deep and meaningful understanding of science content is the ability to engage in modeling.[22,23,24,25,26] Models come in many forms (conceptual, physical, mathematical, diagrammatic, graphical, *et al.*), but all share some basic elements. Students must learn to choose which aspects of a system to model, to choose appropriate representations of these aspects, to make predictions based on their model, and to determine the limitations of their models (where and why a model breaks down).

### B. Experimental Design / Protocol Development

The ability to design and carry out an experiment is a hallmark of scientific thinking, whether or not one becomes an experimental scientist.[27] In the biomedical sciences, experimental design is often referred to as *protocol development* and is a highly prized skill both for future biologists and for medical researchers.[1,3] Yet the skills required to design an experiment/protocol are often ignored in lab curricula across the sciences, leaving undergraduate science majors with no specific training in these skills. By presenting students with a question to answer and minimal guidance (mostly in the form of how to use the high-technology equipment), these open-ended labs encourage students to build these crucial skills.

### C. Error Analysis

As Kung, one of the developers of the SCL reform labs, notes: "Many laboratory courses teach students the mathematics of uncertainty analysis such as the arithmetic mean, standard deviation, and percent error, but students are rarely able to use these constructs to make a strong argument from their data. Even worse, using such tools without understanding may be detrimental to future development of understanding."[14] Thus we have made an effort to raise issues of error analysis in contexts that clearly demonstrate how this skill can help choose amongst competing models of a phenomenon and also translate the challenges of measurement into meaningful interpretation of results.

### D. Technical Lab Skills

Alongside the epistemological framework we present to our students, it is important that they gain practical, technical skills that they will be able to apply in professional contexts. To this end, half of our labs employ an inverted microscope with CCD camera for collecting high-resolution, real-time videos of microscale phenomena. Analysis of such data-rich videos requires an image analysis software, such as NIH's ImageJ (used widely in biomedical research labs), and sophisticated facility with spreadsheet programs. Students are given explicit, extensive support on the acquisition and application of these technical skills, leaving them more time and energy to devote to the open-ended experimental design and data interpretation (for which very few protocols are provided). The other significant high-tech equipment is a USB Infrared to UV spectrometer that collects and displays data in real-time, which is used in labs 10 and 11 in the second semester (see appendix).

### E. Interdisciplinary Thinking

As with other aspects of the NEXUS/Physics project, interdisciplinary thinking is an important aspect of these labs.[7] We aim to foster this connection by carefully choosing laboratory topics and measurement contexts. We also reinforce this thinking by connecting labs where the focus is on physical systems and measurements (e.g. for random and directed motion of beads in lab 3, described in appendix) with labs where physics is measured directly on living systems (e.g. measurements of organelle motion in cells in lab 5, reviewed in depth below). With this scaffolding, we are in a position to ask students to consider what biology can be learned from a physical measurement. They are also asked to seek the biological authenticity in each scenario and to recall, reconsider, and revise their understanding of science content learned in previous courses. It is this thread, together with the specific high-tech tools and modern analysis methods, that most distinguish this new curriculum from the traditional and reform lab curricula.

## III. DETAILED DESCRIPTION OF A NEXUS/PHYSICS LAB

Here we give a detailed illustration of one example laboratory, showing how we have tried to accomplish our goals. A brief introduction to all of the other laboratories for both semesters can be found in the Appendix. As can be seen in the sample laboratory, the skills learned in one lab are applied in other labs, both in new contexts and to encourage interdisciplinary transfer.

**Lab 5, Fall Semester:**
Motion and Work in living systems

**How much work is involved in Active Transport? Classifying Motion and Examining Work in Onion Cells.**

In this two-week lab, which is the capstone lab for the Fall semester, students examine the intracellular motion of organelles, small functional compartments that are visible inside living cells. The example we have chosen is one of the simplest living systems: a thin layer of freshly cut onion skin is actually alive, with many microscopic organelles visible under magnification (Fig. 1a). In order to examine the motion inside of cells, students must be able to successfully operate the microscope with high precision. Students design an experiment to examine the motion of organelles in the cell and consider the different types of transport taking place. Building on tracking skills learned early in the first semester, they make choices as to which areas to track, how many organelles to track, and how to analyze their tracking data. The student's choices and experimental design affect their ability to draw strong conclusions from the data. Ideally, students will track a large number of organelles from a variety of parts of the cell over a long period of time; this gives them a large amount of data from which to draw conclusions about the motion of vesicles inside the cells. This management of large sets of data has been developed throughout previous labs, first introduced in Lab 1, Part 2 (see appendix).

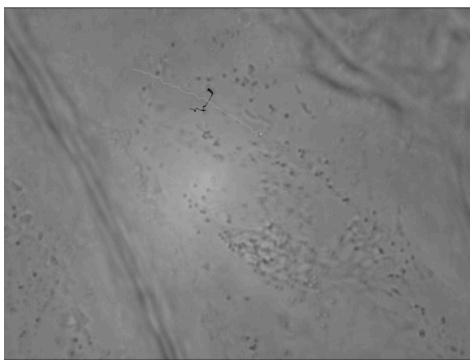

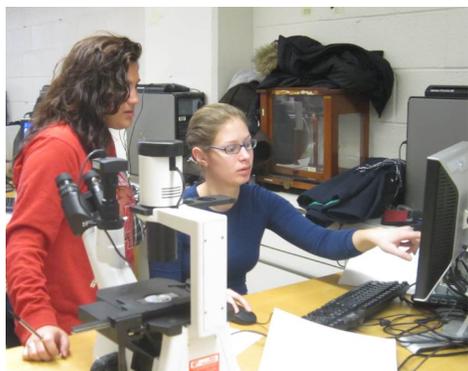

Figure 1: Imaging Organelles in an Onion Skin. a) Image of organelles, small dots, in an onion cell (above)—two of these organelles have been tracked over the course of the video, showing light and dark paths; b) Students working with a microscope to image the onion cell (below).

The goal is for students to examine trends in the data they collect and use them to relate to different models of motion. By analyzing the scaling relations between the distances organelles travel in known times, students can determine if they are moving in a directed manner, in a random manner, or are confined (trapped) in some way.[28] As they learned and practiced in labs 3 and 4 (see appendix), students use a $\log(r^2)$ vs. $\log(t)$ plot of the type of motion that is observed (Fig. 2) to draw conclusions about the method of cellular transport being utilized in that area. Students are asked to connect the quantitative physical analysis they have done with the underlying biological processes.

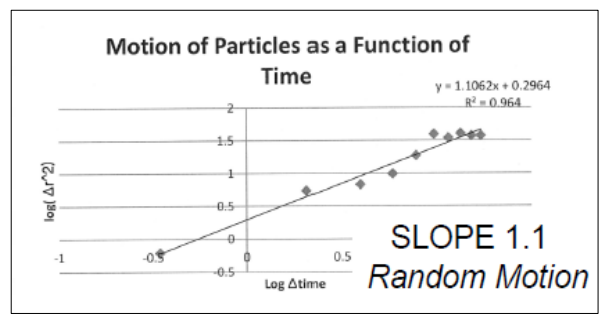

Figure 2: Student plot of $\log(r^2)$ vs. $\log(t)$. The slope of ~1 for the line of best fit indicates random motion.

Finally, the students connect motion to energy and further explore biological implications of the physical measurements. From the work needed to achieve the observed directed motion in the cell, students estimate the rate of ATP hydrolysis involved in the process. This allows them to further realize how the behavior observed inside the cell reveals the underlying biological phenomena at work.

## IV. STUDENT RESPONSE

These labs have been run with two small IPLS test classes at the University of Maryland, College Park, in the 2012-2013 academic year, with the graduate student lab developers (Moore and Giannini) as the teaching assistants (TAs). In the coming academic year, these labs will be run with large enrollment courses, currently scheduled for 240 students, and TAs newly-exposed to the lab curriculum. Throughout the course, multiple assessments were administered to the students to gauge their reaction to the new lab curriculum and to gather information needed to revise the lab curriculum. Though the data gathered thus far represents only a small sample group (maximum N = 31), we have some promising results.

### A. Comparison of NEXUS/Physics labs to ISLE and SCL reform labs: Open-ended prompt

In an open-ended response survey question, researchers for all three curricula (Rutgers' ISLE labs, UMD's SCL, and our NEXUS/Physics labs) asked students to list what they had learned from the labs. Due to the open-ended nature of such a question, students respond with a wide variety of things learned (e.g., physics content, lab equipment used, software programs used) in addition to the more abstract lab skill goals, such as experimental design and analysis and interpretation of data. The frequency of occurrence of abstract lab skill goals can be taken as an indication of student orientation to the value and effectiveness of the curriculum in achieving such goals. For the goals of learning to design experiments and learning to analyze and interpret data, common to all three curricula, we compared data from SCL[18] and ISLE[21] (both N ~ 200, student demographics comparable to NEXUS/Physics test classes at UMD) measured at the end of the respective courses to our students (max N = 31, measured halfway through the 1st (Fall) semester and at the end of the 2nd (Spring) semester). The results, summarized in Table 1, give us hope that we are headed in the right direction. It remains to be seen whether we can replicate these results with a larger course enrollment and different TAs.

Table 1: Student Response to open-ended "What have you learned from these labs?" prompt. Comparison of SCL, ISLE, and NEXUS/Physics lab curricula.

| Learning Goal | SCL (2003) | ISLE (2005) | NEXUS: Middle of 1st Semester | NEXUS: End of 2nd Semester |
|---|---|---|---|---|
| Number of Students | ~200 | ~200 | 22 | 31 |
| Design Experiments | 14% | 24% | 14% | 35% |
| Analyze and Interpret Data | 8% | 11% | 24% | 29% |

### B. Comparison of ISLE and NEXUS/Physics labs: Likert-style survey of goals

Etkina and Murthy employed a student perception diagnostic tool at the end of the ISLE curriculum in 2005 that we decided to use with our students.[21] This diagnostic tool aims to determine if the goals valued by instructors and curriculum-developers are perceived and valued by the students, and also to gauge students' opinions of the effectiveness of the lab curriculum in achieving these goals. Students' perceptions are important because of the strong interaction between expectation and motivation in conceptual change, especially for complex concepts.[21,29,30] Students were asked to consider a list of learning goals that an introductory physics course could have:

- Learn to design your own experiment,
- Learn to interpret experimental data,
- Understand concepts better ,
- Learn to work with other people,
- Learn to communicate ideas in different ways, and
- Prepare for your future professional career;

and were then asked two questions: How important is each goal for you? and How successful were the labs in terms of achieving each goal? Students gave their

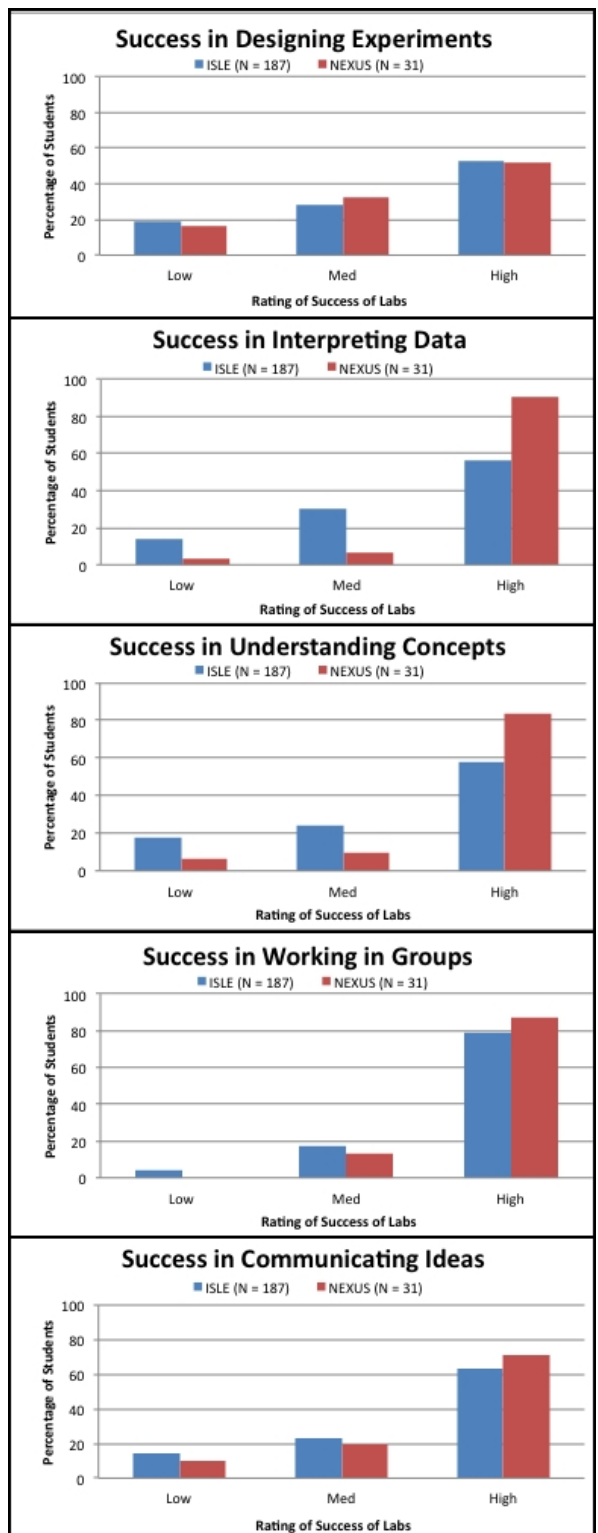

Fig. 3: Comparison between student rating of success of ISLE and NEXUS labs in achieving goals.

responses on a scale from 1 (Not Important/Not Successful) to 5 (Very Important/Very Successful). Fol-

lowing Etkina and Murthy's lead, a response of 1 or 2 was classified as "Low", a response of 3 was classified as "Medium," and a response of 4 or 5 was classified as "High." The relative importance placed on each goal by the two student populations (ISLE and NEXUS) were very similar for all six goals. For the first five goals on the list, the level of importance was also strongly correlated with the level of success; thus, this importance data will not be displayed for the first five goals. Instead, Figure 3 shows a comparison of the students' ratings for how successful the lab curriculum was in achieving each goal. Our NEXUS/Physics labs show comparable levels of success to the ISLE labs.

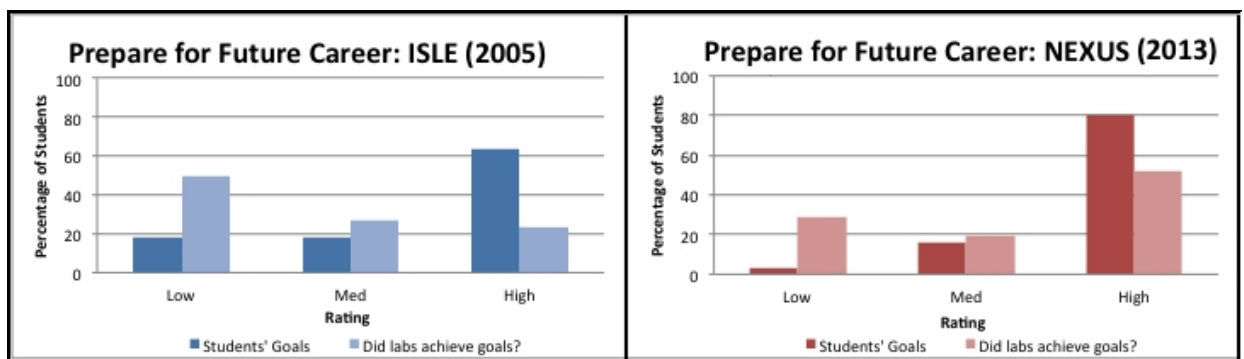

Fig. 4: Comparison of ISLE and NEXUS labs in Importance to Student and Success of Labs in Preparing for Future Career.

The one area where Etkina and Murthy noted a troublesome mismatch between levels of importance and success in the student response was for the goal of preparing for a future professional career. Though their students placed medium to high value on this goal students felt that the labs had medium to low success in achieving this goal. Etkina and Murthy provide plausible explanations for this mismatch, including that students may not be aware of what skills they will need in their future career. The new lab curriculum we have created shows a closer match between student perceptions of importance and success in achieving this goal as highlighted in Fig. 4.

### C. Other student results for NEXUS/Physics labs

In the many surveys used to gain student feedback on the labs (max N = 31), we gathered both encouraging and discouraging comments. As with most new curricula, the students expressed a resistance to change and took a while to adapt to this new way of engaging in learning.[8,31,32] When asked to respond to the open-ended prompt "Are you learning lab skills relevant to biology?" in the middle of the Fall semester and at the end of the Spring semester, approximately 2/3 of students in each survey indicated an affirmative response. While some students remained neutral throughout the course, by the end of the Spring semester there were no students responding in the negative. One student remarked:

> Critical thinking and communication between peers are definitely valuable skills in the realm of biology. Using tracking programs, Excel, and also using microscopes are definitely beneficial skills to have in biology.

When asked if the labs were interesting, another student remarked:

> Yes! Labs such as the neuron lab which relate to background knowledge we already know are especially interesting because we are exploring and proving an area of a subject we did not completely understand!

Students were also asked to respond to an open-ended prompt regarding the experimental design/protocol development thread. They were asked to discuss whether this emphasis was a beneficial or a hindrance, and to describe the advantages and disadvantages of having to create their own protocol. This prompt was administered halfway through the Fall semester and at the end of the Spring semester. The percentage of students finding this emphasis beneficial increased from 64% to 86%. Most strikingly, there were no students in either survey indicating that this emphasis was purely a hindrance. The disadvantage mentioned most frequently by the students was that designing their own experiment takes time and can cause the lab to feel rushed. Advantages noted by the students are illustrated in these two representative comments by different students:

> I find that I am paying attention to every step and can explain the experiment from beginning to end, with an understanding of why things happened and how the results change when a variable is manipulated.

and

> *I definitely see it as beneficial because a large part of science and especially biology is how to create an experiment so this is great practice for the future.*

These results affirm our belief that these labs are headed in the right direction and that future iterations of this new lab curriculum may have similar success, even in large-enrollment courses.

## V. CONCLUSION

The promising preliminary response to our new IPLS laboratories with reformed content and pedagogy highlight that it is possible to sustain the successes of reform pedagogy with a significant topical change, modern equipment, and an interdisciplinary angle. The significant changes to the lab material and needed equipment are guided by the topical shift of the course, but are more focused on the physics on micro- and nanoscales. By explicitly addressing situations where students cannot draw from everyday experience for foothold principles, such as random motion in the first semester, these labs guide students to confront and analyze concepts that are not intuitive.

One significant accomplishment of these pilot labs is the shift of student perception. A majority of our mostly pre-med students considered the physics laboratories as valuable for their future career. We feel that the explicit focus on interdisciplinary connections played a role in this outcome, and hope to gain significantly more insights into this attitude shift when the revised labs will be implemented for large enrollment classes.

## ACKNOWLEDGMENTS

The authors would like to thank the UMD Physics Education and Biology Education research groups, and well as the UMD Physics Department, the Biophysics Program, and the College of Mathematical and Natural Sciences, for their support and feedback throughout this development process. Special thanks are extended to Mark Reeves (George Washington University, Physics), Ben Geller (UMD, PERG), Sergei Sukharev (UMD, Biology), Eric Anderson (UMBC, Physics), Lili Cui (UMBC, Physics), Catherine Crouch (Swarthmore College, Physics), and Karen Carleton (UMD, Biology), for ideas, suggestions, and (where possible) springboard materials. The material presented in this paper is based upon work supported by the Howard Hughes Medical Institute NEXUS grant, and the US National Science Foundation under Award DUE 11-22818. Any opinions, findings, and conclusions or recommendations expressed in this publication are those of the authors and do not necessarily reflect the views of HHMI or the National Science Foundation.

## APPENDIX: BRIEF DESCRIPTIONS OF NEXUS/PHYSICS LABS

**NEXUS/Physics Labs (Fall Semester): [11 weeks]**

**Lab1:** Quantifying motion from Images and Videos.
 **Part 1: How do you quantify motion? Excel Analysis of the 1-D Motion of an Amoeba.** In this one-week lab, students explore the concepts of position, displacement, velocity and acceleration while learning basic Excel skills for data presentation and analysis. The students are given a stop-motion sequence of outlines of an amoeba undergoing roughly linear motion. The students must determine what should be measured, how to measure it, and how to manipulate the resulting data to make statements about the amoeba's motion.
 **Part 2: Can you learn any biology from physical measurements? Analysis of Cell Motion Using ImageJ.** In this one-week lab, students are introduced to digital image analysis. We utilize ImageJ, a freeware program widely used in biomedical research, to analyze short video clips of biological motion. We use the question "For a simple cut, should a doctor prescribe antibiotics?" to open a discussion of the biological context of this lab: the immunological response of a wound to bacterial infections. Students analyze videos of wound-closure, bacterial motion, and neutrophils (white blood cells) to determine their relative speeds. The surprising result (that bacteria are ~30X faster than neutrophils) prompts students to think about both the validity and the implications of their numerical results. The students re-examine their underlying assumptions about how neutrophils operate and hypothesize what mechanisms enable the human body to defend itself in the absence of antibiotics.

**Lab 2:** Inferring force characteristics from motion analysis.
 **How can information about forces be derived from a video? Introduction to Video Capture & Analysis of Directed Motion and Resistive Forces.** In this two-week lab, students examine how various resistive forces scale with the speed of a macroscopic object. Students examine both viscous resistive forces (laminar flow) and drag forces (turbulent flow). Students capture and analyze their own video samples for coffee filters falling through air and for plastic and metal spheres falling through various concentrations of glycerol solution. The students begin examining error analysis and how experimental design and data collection decisions affect the uncertainty in an experimental

result. We use macroscopic objects to allow students to tap into their intuitive physical experiences and to begin the process of connecting their everyday experiences with physical laws. Students are introduced to the process of (1) representing observations in multiple ways (graphs, equations, diagrams) and (2) using these multiple representations to make statements about a physical model (e.g. via fitting or via scaling arguments)—a process that is a core component of the rest of the labs.

**Lab 3:** Observing Brownian motion at a microscopic scale.[33]

**What does 'Random' motion look like? Describing Diffusion and Random Motion.** In this three-week lab, students explore the nature of random motion generated by the rapidly varying (i.e. random) forces exerted by thermal fluctuations. Brownian motion is ubiquitous in living systems at a molecular and cellular level and is distinct from the types of motion in students' everyday experiences. They look at how diffusive motion scales with time for microspheres (beads) suspended in solution. Using the microscopes, students gather their own videos and use ImageJ to collect data on the positions of the beads. They then make histograms of the motion at various times. Throughout this lab, each student group has been working with intentionally varied materials: different bead masses, bead sizes, and fluid viscosities, such that each lab group has one changing parameter and two controlled parameters. Finally, using the collective data produced by all the groups, students determine which factors (mass, size, or viscosity) affect the diffusion constant for random motion and determine a mathematical model for how the diffusion constant will change with each parameter. Building on their experience in Lab 2, scaling and modeling arguments are used extensively to make sense of physical observations.

**Lab 4:** The competition between Brownian motion and directed forces.[34]

**How large a force is needed to transition from random to directed motion? Random vs. Directed Motion.** Machines on microscopic scales, such as molecular motors or ion pumps in living systems, constantly compete against and take advantage of the thermal forces that create random Brownian motion. In this two-week lab, students use beads of different sizes suspended in solution to explore the crossover from random to directed motion as an external force is applied. By tilting the microscope and thus inclining the slide, students can gather and analyze videos to determine the motion of the different sized beads under a condition where the gravitational force is affecting bead motion in addition to thermal forces. Students can directly observe the crossover from random motion where thermal forces dominate to directed motion where gravity dominates and analyze how it depends on the size of the bead. Due to differences in how displacements scale with time, even small directed forces lead to directed motion on long enough timescales; thus students also examine the effect of the time frame over which the bead is followed.

**Lab 5:** Motion and Work in living systems.
**How much work is involved in Active Transport? Classifying Motion and Examining Work in Onion Cells.** (See example lab description, Section III.)

**NEXUS/Physics Labs (Spring Semester): [11 weeks]**

**Lab 6:** Modeling fluid flow.
**How do channel geometries and arrangements affect fluid flow? Exploring Fluid Dynamics and the Hagen-Poiseuille (H-P) Equation.** In this two-week lab, students examine the relations between pressure, flow rate, and pipe-width for a fluid containing microspheres flowing through channels arranged in parallel or in series. Students model the expected results using the H-P equation and then develop an experimental protocol to test their predictions. Students discuss connections of their observations to biological fluid channel networks (e.g. the circulatory system) and to the design of health interventions (e.g. cardiac bypass).

**Lab 7:** Analyzing electric forces in a fluid.
**How do charged objects in a fluid interact with each other and respond to external electric fields? Electrophoresis and Charge Screening in Fluids.** In this two-week lab, students explore how microspheres that are suspended in solution move in an electric field. Students discuss that beads charge spontaneously when placed in solution. By varying fluid salinity, the effective charge is changed due to screening of charged objects by the surrounding ions (Debye screening). Using the basic vocabulary of colloidal chemistry, students model the mechanism of electrophoresis, examining how the terminal velocity of the bead's motion in an electric field is related to the effective charge of the bead.

**Lab 8:** Modeling signal transmission along nerve axons.
**What affects the distance over which an electrical signal is transmitted and the speed of transmission? Testing Models of Signal Transmission Along Nerve Axons.** In this two-week lab, students model signal transmission along nerve axons and explore the mechanisms by which signal speed and the distance

traveled by the signal can be increased. Students explore conceptual, mathematical, diagrammatic, and graphical models for signal transmission in order to design and test physical models of electrical circuits using resistors, a breadboard, and voltage probes with the aid of Logger Pro software. Students explore what biological insights can be gained from the measurements on these physical models—in particular, insights about the role of myelination and issues related to nerve malformation. (Adapted from Eric Anderson and Lili Cui at UMBC and Catherine Crouch at Swarthmore College.)

**Lab 9:** Introducing geometric optics through experimental observations.[35]

**How can microscopes magnify objects? Exploring Light and Lenses.** In this two-week lab, students explore basic optics components and principles as a first exposure to light and lenses using a vertically positioned optical rail. While the study of magnification in traditional geometric optics labs involves the movement of an object and analysis of the corresponding image position, this is not reflective of the optics systems in microscopes and other biomedical imaging devices. We designed this optics lab with some realistic constraints found in real microscopes, such as a fixed total optical length, which we enforce via the vertical arrangement of the optics. Students collect information about the focal lengths, as well as object and image distances. This data sets the stage for the core lab activity, where students develop and assess their own mathematical models relating the image and object distances to the focal lengths. Students discover that even for a fixed total optical length, each lens can form a clear image by adjusting the lens position, and that each resulting image has a different magnification.

**Lab 10:** Analyzing light spectra and exploring implications for living systems.[36]

**How can measurements of light spectra provide insights into the nature of matter and the characteristics of living systems? Spectroscopy—Exploring Emission, Absorption & Evolutionary Adaptation.** In this two-week lab, students use a variety of spectroscopes (rudimentary to high-tech) to explore emission and absorption of light. Examining the Hydrogen spectrum, students investigate the Bohr model of the atom—in particular the electron level transitions that result in the Lyman, Balmer, and Paschen series. Students then explore how different combinations of sources and filters affect observed spectra, including everyday filters such as water or sunglasses. Using this accumulated knowledge, students hypothesize which environmental factors may have constrained the evolution of the spectrum that is visible to the human eye and gather evidence to support the plausibility of their hypotheses.

**Lab 11:** Exploring complex absorption and emission in molecules.

**Must 'what goes in' be the same as 'what comes out'? Spectroscopy & Fluorescence in Chlorophyll.** In this one-week lab, students examine the fluorescence of a chlorophyll solution when exposed to a variety of light sources in order to gain a foundational understanding of the concepts of how light is absorbed and emitted by fluorescent molecules. Students then consider the physical implications of the ways in which fluorescence is presented in other science venues, such as fluorophore excitation and emission spectra.